# Autonomous CPS mobility securely designed

David Hofbauer[1], Christoph Schmittner[2], Manuela Brandstetter[3], Markus Tauber[3]

[1]*Forschung Burgenland, Eisenstadt, Austria*
[2]*Austrian Institute of Technology, Vienna, Austria*
[3]*University of Applied Sciences Burgenland, Eisenstadt, Austria*

*Abstract*—In the last years the interconnection and ongoing development of physical systems combined with cyber resources has led to increasing automation. Through this progress in technology, autonomous vehicles, especially autonomous trains are getting more attention from industry and are already under test. The use of autonomous trains is known for increasing operation efficiency and reduction of personnel and infrastructure costs, which is mostly considered for main tracks. However, for less-used secondary lines, autonomous trains and their underlying sensor infrastructure are not yet considered. Thus, a system needs to be developed, which is less expensive for installation and operation of these trains and underlying infrastructure for secondary lines. Therefore, this position paper describes the process of how to derive an approach to help develop a digital interlocking system at design time for the use with secondary railway lines. In this work, we motivate the necessary research by investigating gaps in existing work as well as presenting a possible solution for this problem, a meta-model. The model considers safety, security as well as interoperability like 5G and socio-technical aspects to provide a holistic modeling approach for the development of the interlocking system for industrial secondary line use cases.

## I. Introduction

Significant progress has been made in the field of automated and autonomous train services in recent years and the first autonomous trains are already under test [1]. One of the main motivations for increasing automation of train traffic is the associated increase in maximum capacity utilization. A digital and communication-based moving block system, in contrast to the classic fixed-block system defined by infrastructure, can enable a significantly higher number of trains without a physical expansion of the track infrastructure [2]. Based on this motivation, known systems, for instance the European Train Control System (ETCS Level 3), are primarily designed for the use on main tracks. Such modern rail control and management systems are currently in delayed or reduced development on secondary railway lines, compared to main lines. Secondary (branch) lines, which are used e.g. in rural areas or also for industrial purposes, are mostly still controlled by classic train protection and control systems. To improve these, in addition to increasing the efficiency of operations, cost savings through a reduction in personnel and infrastructure costs are a potential advantage of autonomous train services for secondary lines. This requires systems that are less expensive to install and to operate. In the urban railway sector, communication-based train control (CBTC) systems show that automation and increase in efficiency is feasible. Such systems take advantage of the restricted and controlled environment by using for example WLAN instead of GSM/LTE based communication [3].

By taking advantage of the lower speeds and train numbers on secondary lines, European Rail Traffic Management System (ERTMS) compatible systems based on commercially available hardware and by utilizing existing communication infrastructure must be developed. When investigating this development of autonomous train services, many interconnected components need to be taken into account and addressed to avoid certain threats and vulnerabilities. This interconnectivity of components in combination with security, safety and interoperability (e.g. concerning external interfaces and socio-technical factors) needs to be considered.

To address these issues, an interlocking system for autonomous trains on secondary lines will be developed. Interlocking systems serve as the basis for autonomous trains and enable the interaction and control with the railway. This will be based on an existing interlocking system for industrial railways, which is already using industrial off the shelf components. This explains the practical use case, which will be developed in the context of the acknowledged future research project. However, to support the development and design of the above-mentioned interlocking system, by considering security, safety as well as interoperability aspects based on actual industrial use cases, a meta-model will be developed. Thus, risk analysis, security & safety concepts as well as external interfaces must be evaluated and adapted. This is also motivated by the potential usage of wireless communication like 5G and when incorporating $3^{rd}$ party and end user devices, for which human socio-technical factors must also be considered. The interlocking system, combined with a newly developed management system, can be used as a basis for further automation towards fully autonomous systems. Particular emphasis is placed on what we call Railway Operation as a Service (ROaaS), a service for managing the sensor data and derive appropriate actions in a logical central and scalable place e.g. a cloud, as well as end user aspects such as human-machine interaction and communication with external systems and their impact on hazards and risks, as well as user behavior in relation to the system.

Therefore, the overall research issues addressed by our work beyond this position paper focus on providing an approach to support the design of a digital interlocking system by integrating security and safety as well as interoperability (external devices, 5G, socio-technical) aspects for the autonomous railway domain. This should help to identify threats and vulnerabilities at design time and thus to build a secure system. Our proposed solution for this approach will be a meta-model for autonomous trains, which includes all the above-mentioned aspects, that should be easy to understand and to reuse for further use cases in this industry. It will include all mentioned aspects by dividing components into different component perspectives (e.g. safety, security, etc.), which will be described through modeling languages in detail. Use cases can then be checked with this model, to see, which components need to be addressed in the design phase of the system. This contribution should help improve operation efficiency and reduction of costs on less-used secondary lines. However, the purpose of this position paper is to investigate the need of this research by providing an outline of gaps in existing work. Further the potential of the meta-model should be described by showing our process of how to develop one. To summarize, the position paper investigates a meta-model based approach for supporting the safe and secure design of such a cyber-physical system (CPS) for controlling and managing autonomous railway systems.

The remainder of the paper is structured as follows: Section 2 shows related work concerning the most important topics of our paper and how our work will enhance and use this existing literature. Next, section 3 will describe the work towards the development of the meta-model. It will discuss information concerning use cases, interoperability, background information about available technology and approaches as well as the actual model derivation. Finally, in section 4 we summarize the approach and how our work beyond this position paper will go on.

## II. Related Work

Related work includes research concerning autonomous vehicles, meta-modeling, consolidation of threat and vulnerability catalogs as well as security, safety and interoperability aspects, which should show the gaps in existing work and how we will enhance it.

Zafar et al [2] examined railway interlocking systems concerning safety. They conducted a formal analysis of safety properties of moving block systems to prevent collision and derailing of trains at critical components of train systems. By using graph theory and Z notation, a step-by-step procedure was done to reduce the complexity of these systems. They formalized the abstract safety properties and afterwards redefined them by applying it to the moving block system. Also, computer-based controls for monitoring trains and state space was done. Our work could investigate the use of this approach for modeling safety aspects. Nevertheless, they did not investigate interoperability and dependencies to other aspects, e.g. security, which we will examine in our work by creating a meta-model, which also includes security, external factors as well as their dependencies.

Strobl et al [4] investigated threats and vulnerabilities regarding autonomous cars. Based on component and data flow modeling, security threats and vulnerabilities concerning each component were gathered by investigating related literature. A catalog was derived, which lists all found threats and vulnerabilities per system component. These findings were in the end used to find the most crucial component of a connected car with hindsight to security. This process of component modeling and the derivation of the catalog can also be used for the development of the meta-model, which also has to include threats that need to be addressed when developing a secure interlocking system. But in contrast to this work we will investigate autonomous trains on secondary lines in combination with safety aspects.

In [5] Schmittner et al. investigated so-called cyber-physical systems (CPS), which combine physical machines with computational components. This correlation posed new challenges in terms of security and safety. Previously, these were treated separately, but through this combination, new holistic approaches for safety and security analyses are needed to identify safety failures and security threats as well as their interoperability. Thus, they investigated two promising methods for this identification through the help of a use case concerning connected, intelligent vehicles. The outcome showed that both used methods have the problem that interoperability between safety and security was integrated insufficiently and needs improvement. Our work will make use of such identification methods but extends them by integrating the needed interoperability in a meta-model for autonomous trains.

Concerning meta-modeling, Bicaku et. al [6] investigated how to ensure trustworthy communication from edge devices to the backend infrastructure in Industry 4.0. Therefore, they developed a meta-model based on existing ones, e.g. RAMI4.0, to describe end-to-end communication for Industry 4.0 use cases. They also addressed dependencies, interfaces and security challenges in their developed model. This outcome can now be used as a tool to identify dependencies between Cyber Physical Production Systems (CPPS) components to help define clear monitoring points for a system. In contrast to this work, our paper focuses on developing a meta-model for autonomous trains. Nevertheless, their approach of deriving a model as well as including dependencies can also help developing our meta-model, especially concerning the security part.

Bless et al. [7] investigated the deployment of critical infrastructure to the cloud caused by its rapid and flexible service deployment possibility. However, this raised questions concerning privacy, security and resilience, which are stricter in this sector than in the general consumer market. Based on legal regulations, they investigated salutations for fault identification and localization. But to understand such a complex system, a meta-model needed to be developed. Thus, they derived an architectural model for critical cloud infrastructure services in cloud computing. Further investigation concerning moving critical services to the cloud was done in [8], in which a risk assessment method for this critical infrastructure shift was presented for which they extended well established information security risk assessment methods. Our work can use and build on this meta-model derivation as well as risk assessment methods because it also concerns critical infrastructure. But in contrast to their work, we will assess risks and develop a model concerning autonomous train secondary line use cases.

In this work [9] an approach was investigated of how to derive requirements and controls from important known security standards. A catalog was formed as a security checklist for secure use cases in the Industry 4.0 domain. One outcome showed, which requirements in an industrial security use case need to be investigated the most caused by its importance and thus shows how controls from standards can be fulfilled to achieve higher security. Our work can make use of this investigation approach for the modeling of autonomous train components to find the ones, which are more important than others. Also, the controls derivation method could be used in our work. But in contrast to this paper, we focus on developing a meta-model and including standards for autonomous trains, which should include next to security also safety standards.

Brandstetter et al. [10] researched major impulses for technological impact assessment in general as well as in specific security matters. In preliminary studies (e. g. about the implementation of steganographic methods in organizations), knowledge-sociological hermeneutics in a participatory design was used to investigate the need for discretion of users, employees, experts, researchers and key decision-makers. They found out, that shared responsibility and trust as well as good information concerning new technologies are the major conditions for a proper function of the technology itself. Via simulating highly sensitive issues and scenarios, the technology evaluation is carried out in this project too. Social and legal impact of the meta-model is directy linked to a solid confidence-building programme. This work can be used as basis for our socio-technological aspects investigation and integration into the meta-model and will thus be enhanced by social aspects concerning autonomous trains.

### III. META-MODEL DEVELOPMENT

In this section, we present our intended research approach to develop a meta-model, which will be divided in three sub-sections: (A) Use Case: Description of the use case, for which the meta-model will be derived; (B) Interoperability: Aspects, which must be investigated and included in the meta-model will be listed in order to understand the meta-model build-up; and (C) Modeling, which presents our first modeling ideas and approaches.

#### A. Use Case

Through the growing interconnection of components within trains and the usage of Internet-of-Things (IoT) technologies and sensors, trains get increasingly complex but also increasingly autonomous. To develop these autonomous trains, which are thus so-called CPS [5], new approaches and tools are needed, which concern the components' dependencies and interoperability (e.g. safety and security). We will develop such a tool, which will support the cost-efficient design and development of a digital train interlocking system.

Our work is based on use cases, which will be investigated in detail and with the help of industry partners, to derive the meta-model. More precisely, we will examine autonomous trains on secondary branch lines. Especially for secondary lines, autonomous trains can achieve higher mobility for e.g. rural areas. Furthermore, autonomous systems can lead to a cost reduction using cheaper new infrastructure (e.g. IoT sensors) and partly already used infrastructure instead of building everything from scratch. The use cases for secondary lines will be identified by the help of industry partners and will in the end be evaluated with the meta-model to identify everything, which needs to be considered in order to design and develop a secure digital interlocking system.

#### B. Interoperability Aspects

To develop reliable and resilient systems, it is important to identify *security and safety* relevant aspects prior to the actual development. This topic is a very complex one in the railway sector. Safety has always been part of reliability availability, maintainability and safety (RAMS), which is regulated within standards [11], [12]. It has been addressed by the use of a safety case regulated by these standards, security elements were only addressed concerning access protection to railway controls systems. Therefore, a holistic approach is needed, which must combine safety as well as security. Security and safety relevant attributes of system components will be an essential part of the meta-model, also considering threats and vulnerabilities for each component. Through the help of certain security and safety standard controls [9] as well as the usage of fundamental safety and security analysis tools (e.g. STRIDE, DREAD, FMEA), components as well as their threats and vulnerabilities can be identified and addressed.

Previous work on meta-models often addressed industrial plants [6], which have overlapping with the railway domain but also distinct differences. Railway systems are geographically much more distributed than industrial plants and instead of building an own communication infrastructure, an *external* common infrastructure can be used (e.g. 5G in the near future). When using wireless communication, another important factor, which needs to be considered concerns external 3$^{rd}$ party components, for instance end-user or IoT devices. For this, also socio-technical questions need to be investigated to get an overall view of the autonomous railway system, which must include every tiny important aspect. Therefore, our meta-model will address this interoperability by integrating, next to security and safety, dependencies of 5G technology and socio-technical aspects as well as concerning end-user devices and 3$^{rd}$ party components connected with the railway system.

In general, *interoperability* is an important issue in the railway domain: on the one hand rolling stock and infrastructure need to be compatible, which is provided by different manufacturers and all have different lifecycles. On the other, infrastructure is widely stationary but rolling stock is in motion. For this, the system must work reliable and all components need to cooperate with each other in a secure way. Thus, our meta-model will include this topic by also investigating railway interoperability frameworks, e.g. [13], but will extend them by integrating 3$^{rd}$ party devices, 5G, socio-technical aspects, security as well as safety. The next section explains in detail the derivation of a meta-model including all these aspects.

#### C. Modeling

An important tool to design and document complex systems is an adaptable meta-model. In the railway domain, the Euro-Interlocking Project of the UIC tried to develop such a model but no official solution has yet been published. Thus, at the moment, interlocking systems are adapted by proprietary projection tools or manually to laws and regulations in different countries, which is one of the main factors for high costs of these systems [14].

Caused by this, meta-models from other industrial domains need to be investigated to find an appropriate modeling solution for the railway domain. These can serve as motivation and basis for the development of the railway meta-model. One promising reference model is the reference architectural model Industry 4.0 (RAMI4.0) [15], which serves as a guidance for Industry 4.0. It describes key elements of objects or assets based on the usage of a layer model, composed of three axes: (i) architectural axis (decomposition of a machine into its properties structured layer by layer), (ii) process axis (life cycle & value streams) and (iii) hierarchical axis (IT & control systems). It also included an "administrative shell", that shows a virtual representation of real assets and provides information about status of assets and data during its lifecycle.

A further meta-model was derived based on RAMI4.0, which should support the understanding of complex systems, cyber-physical production systems (CPPS). It was used to describe autonomous systems and is an example usage of RAMI4.0. This CPPS meta-model can thus be used as a tool to describe different CPPS use cases for Industry 4.0 [6].

Another example is the Internet Reference Architecture (IIRA) [16], which is an open standard architecture for the design of Industrial Internet Systems (IIS). Based on ISO/IEC 42010 standard specifications for complex systems consisting of multiple components and interconnected systems, the IIRA defines the most important components of the industrial internet architecture categorized in: Implementation, function, usage and business viewpoint.

To represent physical things in the digital world, to model and describe them, a so-called "Digital Twin" can be used. A "Digital Twin" consists of a physical product, a virtual product and connected data that tie the physical and virtual product [17]. It can be seen as a "shell" that, depending on the application, further consists of sub-models. There are already sub-models that concern security and safety as well as depending standards but they are considered as static information and do not concern interoperability or autonomous vehicles.

The above-mentioned meta-models are a good basis for our planned modeling approach but are too general for the concrete use case. Also, currently there are many development approaches concerning the railway domain, which offer a broad perspective: (i) The RailTopoModel [18] offers a common language for railway infrastructure data, which could be use in every area of the railway domain; (ii) Planpro [19], developed by the "Deutsche Bahn", was developed to support the planning and construction of control systems in general and electronic interlocking systems; (iii) the so-called infrastructure data management for transportation companies (IDMVU) [20] serves as the basis for a logical data model based on infrastructure assets for transportation companies; (iv) lastly, the outcome of the international project IFC-RAIL [21] will provide a global standard for the planning, drafting, constructing as well as maintenance for railway systems. But none of these mentioned approaches concern the digitalization of interlocking systems for secondary lines or a holistic approach by including e.g. safety, security and other mentioned aspects and thus can just be used as the basis for our meta-model for autonomous trains.

However, for the design and documentation of use cases in the autonomous railway domain, it is important to provide such a structured modeling tool for the definition and explanation of all components. Especially, to comprehend and document the dependencies between security and safety aspects of components, the meta-model is necessary. Furthermore, interoperability concerning new communication technologies (5G) as well as corresponding 3rd party devices and thus socio-technical aspects needs to be addressed in this domain. For this, typical components concerning automated railway control systems will be consolidated, cataloged and its dependencies will be integrated in a meta-model. The outcome should support the development of a secure digital interlocking system at design time.

To create a meta-model, a specific modeling language must be used. UML (unified modeling language) or SysML (systems modeling language) are known languages. Since UML was adopted, it became more and more popular for software engineers [22]. This software focus has discouraged systems engineers at first but has been adapted to the systems engineer domain over the years. This led to the development of SysML, a standard modeling language for systems engineering, which should support the modeling of a broad range of systems through adapting UML and adding better terminology. Nevertheless, both modeling languages are appropriate for the autonomous railway meta-model and will be used. For the creation of the meta-model itself, ADOxx [23] or the Eclipse Modeling Framework [24] are possible modeling software solutions.

Figure 1 shows an example of how a possible meta-model could look like. In general, a meta-model is based on perspectives [25], which for our use case, designing a digital interlocking system for secondary lines, could be e.g. safety, security, social as well as needed hardware and software. Within perspectives, components can be added, for example hardware components can be the mentioned external devices or certain cables, which connect components, whereas software can be a specific app to control hardware or to monitor certain safety aspects. Security could be a protocol or an encryption algorithm, socio technical aspects could include passengers (using mobile phones) and safety could include e.g. the locking of train doors. All autonomous railway components could easily be integrated. Dependencies and interoperability are shown by so-called connectors, which connect every necessary component and thus perspective with each other, which is a goal of our meta-model. This approach of Figure 1 only shows one possible progress and buildup of the meta-model and is not a final version. However, it should give a first idea of how it will look like. The aspects within each perspective (Device, App, etc.) are just examples. All real aspects will be listed, described and evaluated in detail in the future meta-model.

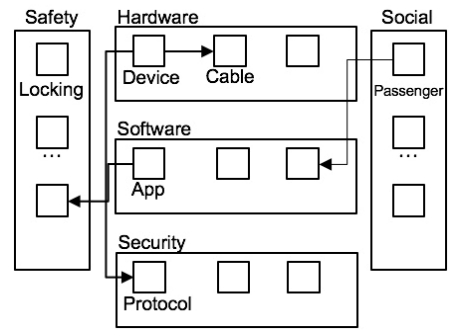

*Figure 1: High level meta-model approach for ROoaS*

The next step would be the description of the model. XML would be a possible solution to achieve this. In a high-level XML code, Figure 1 could look like this:

*<Class>*
  *ID*
  *Safety < Component 1, C2, Cn >*
  *Security < C1, C2, Cn >*
  *Social < C1, C2, Cn >*
  *Hardware < C1, C2, Cn >*
  *Software < C1, C2, Cn >*
*</Class>*

*<C1>*
  *CID*
  *Type< CloudSec, Tracks, IoT, 3rdParty >*
  *Threat<T1, T2, Tn>*
  *Vulnerability<V1, V2, Vn>*
  *Dependency<D1, D2, Dn>*
*</C1>*

The class (*<Class>*) describes the modeling approach in general. All perspectives (in this case: safety, security, social, hardware, software) are included and within perspectives the components are described. To explain further, an example subclass of a component is shown (<*C1*>), which has to include the component's ID, the type (e.g. Track, IoT Devices, etc.), threats, vulnerabilities as well as dependencies to other components. By using XML, every component, perspective, threat, dependency, etc. can be described in their own subclass, which creates a simple yet hierarchical build-up structure. Figure 2 thus shows this structure with safety as example tree. Again, as explained above, this is just a first idea to show a possible approach and will be evaluated in detail in our future work beyond this position paper.

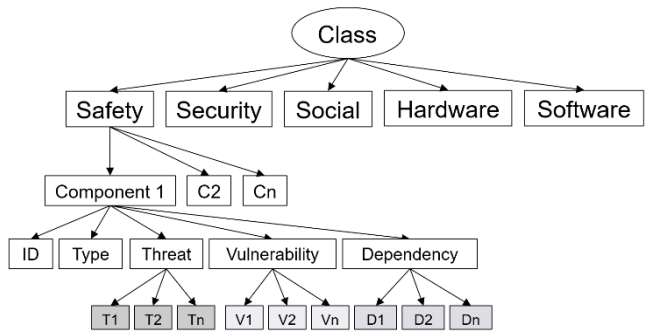

*Figure 2: Hierarchical structure of possible modeling approach*

Through such a simple approach, the build-up of a meta-model would be well-described and easy for everyone to understand. This is very important to support the design of the digital interlocking system as well as to reuse the meta-model for future autonomous railway domain purposes.

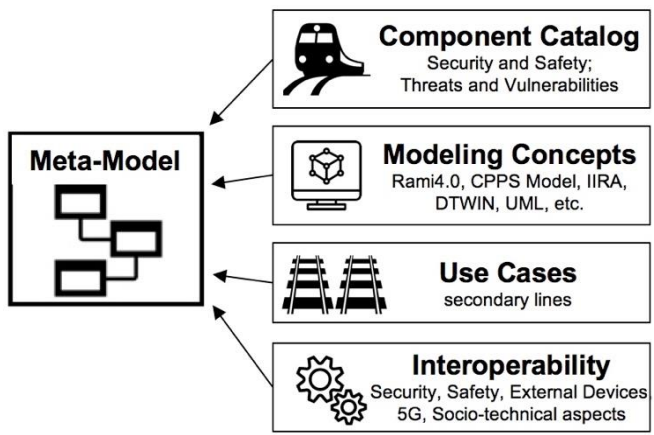

*Figure 3: Overview of the meta-model components*

Figure 3 shows a final overview concerning this modeling section. The meta-model will be based on these four "pillars", which will all influence the setup of the model and thus summarize our approach:

- Component Catalog: This catalog will be used to build the meta-model from scratch. Security as well as safety related standards, threat and vulnerability catalogs, etc. concerning autonomous trains need to be investigated and developed to get an overview of all components that need to be secured.
- Modeling Concepts: Caused by the fact that no usable meta-models concerning our topic have been produced until now, we have to investigate already in use models and approaches (also from other domains) to get ideas and develop a steady meta-model concerning all components.
- Use Case: A use case for autonomous trains on secondary (branch) lines will be investigated. The potential advantage of this is the increasing efficiency of operations as well as cost savings through reduction in personnel and infrastructure.
- Interoperability: A main aspect, which enhances our work in contrast to others is the interoperability factor. Often this is not addressed when working with topics like security and safety. In addition, we will include and investigate these in combination with new possible common communication channels (5G) and thus socio-technical aspects concerning 3[rd] party devices (e.g. end-user devices, IoT).

## IV. CONCLUSION & FUTURE WORK

In this position paper, a meta-model based approach for supporting the safe and secure design of cps for controlling and managing autonomous railwail system is investigated. A use case in the railway domain was presented. Less-used secondary (branch) railway lines can benefit from a current topic, the autonomous railway. More and more lines are closed because operation costs cannot be covered but through the help of such cyber-physical systems, autonomous vehicles, higher efficiency of operations and cost reduction through personnel reduction can be achieved. For this, less expensive systems to install and operate are needed and new approaches have to be developed. Thus, a digital interlocking system for autonomous trains on secondary lines, which we call Railway operation as a Service, should be developed. But to accordingly design and build this, a tool is needed to support this development already at design time. A possible way to do this, is to create a meta-model, which can help model the use case and provide a holistic approach of how the system should look like and to show, which important aspects need to be considered.

When looking at related work, separate topics concerning this system have been investigated but none of these approaches have been combined or adjusted to the railway domain: Digital interlocking systems have been designed but not adopted to autonomous trains and secondary lines; many meta-models were produced but not for the railway domain for autonomous trains for this use case; security and safety were often investigated but never efficiently combined for this domain in an appropriate way. Although none of this related research has addressed our topic appropriately, all these approaches have to be investigated to see if they could also fit for the development of our meta-model.

Therefore, the future meta-model for the use case should include many specific aspects as well as their interoperability and dependencies. By creating a component catalog and adding threats and vulnerabilities to it, an overview of the railway should be defined. By using already established meta-model approaches from other domains as basis ideas, a new meta-model should then be formed by also integrating the catalog. The model should especially be focused on the following aspects: Security, safety and interoperability, which includes new communication channels for the autonomous train (5G technology) as well as 3[rd] party devices (mobile devices, IoT) for which also socio-technical dependencies will be investigated and included.

The approach presented in this position paper should be a first possible projection of the meta-model derivation.

Therefore, a setup of perspectives (Figure 1), which include the focus aspects, is presented for the meta-model. This can be easily described in XML and can also be modeled by using notations e.g. UML or SysML. Through this setup, a hierarchical structure could be derived (Figure 2), which has the advantage of being easy to understand and to reproduce, which is important for possible further use cases in this domain. Figure 3 shows the meta-model approach again in general, which includes every important aspect that needs to be considered during the development of the model.

Thus, the current situation of the secondary line use case will be improved by the development of this meta-model. No approach concerning this use case and all mentioned aspects has been developed or used until now. Therefore, a holistic, overall view specifically addressing this sector will be provided by our future meta-model.

The future work beyond this position paper comprises the development of the meta-model with all its interoperability aspects. The next steps include the exact definition of the use case in conjunction with a requirements catalog. Based on that, components should be derived, cataloged and threats as well as vulnerabilities gathered. During this phase the safety, security and interoperability aspects for the use case and meta-model are also investigated and integrated. After finishing the meta-model, the applicability of it should be evaluated in a real scenario, where the development of a digital interlocking system for Railway operation as a Service should be supported at design time using the meta-model.

ACKNOWLEDGMENT

Research leading to these results has received funding from the project BESTE-AB, funded by the Austrian Research Promotion Agency (FFG), coordinated by the Austrian Institute of Technology.